\documentstyle[aps,prl,twocolumn,psfig]{revtex}

\begin{document}
\draft
\author{John F. Dobson $^{1,2}$, Jun Wang$^1$ and Tim Gould$^1$}
\title{Correlation energies of inhomogeneous many-electron systems}
\address{$^{1}$Research Centre for Modelling and Computation, School of Science,
 Griffith University, Nathan, Queensland 4111, Australia}
\address{$^{2}$Groupe de Physique des Solides, Universit$\acute{e}$ de Paris, 2 Place Jussieu, 75251 Paris, France}
\date{November 23, 2001}
\maketitle

\begin{abstract}
We generalize the uniform-gas correlation energy formalism of Singwi, Tosi,
Land and Sj\"{o}lander to the case of an arbitrary inhomogeneous
many-particle system. For jellium slabs of finite thickness with a
self-consistent LDA groundstate Kohn-Sham potential as input, our numerical
results for the correlation energy agree well with diffusion Monte Carlo
results. For a helium atom we also obtain a good correlation energy.
\end{abstract}


Despite eighty years of study, the accurate calculation of the correlation
energy of interacting quantal many-electron systems is still often a
challenge, even for some systems not regarded as ``strongly correlated''.
For realistic many-electrons systems the current state of the art includes
diffusion/Green function quantum Monte Carlo (DMC), variational and
quantum chemical methods such as the Configuration Interaction (CI) approach.
These all have stringent practical limitations to relatively small and/or
not-too-complex systems. Approximate density functionals of the
local-density (LDA) and generalized gradient (GGA) class \cite
{KohnBeckeParrChemDFTReview96} are in principle less accurate than the above
approaches, but they remain feasible even for very large, complex systems,
and often provide useful accuracy. They fail completely, however, to
describe long-ranged correlations in cases where these differ qualitatively
from those of the homogeneous electron gas. A case in point is the van der
Waals (vdW) or dispersion interaction:\ all LDA/GGA approaches miss its
long-ranged part completely\cite{KohnMeirMakarovVdWPRL98}\cite
{DobsonWangPRL99}\cite{JFDEtAlAustJChem02}, and give at best patchy results
at short range\cite{RareGasDimersPattonPederson},\cite
{PW91ForVdWHBondTsuzuki}. Thus DMC, CI\ and standard DFT methods are all
likely to be problematical for large complex vdW systems of practical
interest, including many soft-matter cases.

Here we present a new general approximation method (``ISTLS'') for the
correlation energy of{\em \ inhomogeneous} electronic systems, which we
believe will be appropriate for vdW problems among others. It employs a
self-consistent scheme for the pair distribution based on the surprisingly
accurate {\em homogeneous}-gas correlation energy method of Singwi, Tosi,
Land and Sjolander (STLS)\cite{STLSOrig}. We take as input an approximate
Kohn-Sham (KS)\ potential $v_{KS}(\vec{r})$ of the inhomogeneous system, and
produce the correlation energy $E_{c}$ as output. Tractable input theories
such as LDA or exchange-only Krieger-Li-Iafrate\cite{KLIPRA93}, by
themselves, make large correlation-energy errors of order +100\% and -100\%
respectively, yet starting from their KS potentials our ISTLS theory yields
correlation energies accurate to a few percent. In its numerical complexity,
and also in its accuracy for $E_{c}\,$of large systems, ISTLS appears to be
intermediate between LDA/GGA and the more microscopic approaches mentioned
above.

Our original motivation for deriving ISTLS was primarily to address
soft-matter problems such as polymer cohesion or the energetics of
graphite and its intercalates. For such vdW systems, the RPA-like\cite
{DobsonWangPRL99} nature of the method, together with its uniquely
self-consistent local- field correction, suggests it will accurately
describe vdW interactions\cite{KohnMeirMakarovVdWPRL98} while also correctly
treating other kinds of bonding\cite{JFDEtAlAustJChem02}. The method is by
no means restricted to vdW problems, however, and should provide a useful
alternative approach both for soft and for hard matter. In particular it
appears to be highly competitive with recently-proposed correlation energy
theories based on the GW method\cite{vdWQuasi2DFromGWGodby}
; it gives a much more accurate correlation
energy in the homgeneous gas at large $r_{s}$, for example. Being
intrinsically approximate, ISTLS needs to be tested. Therefore, in the
present paper, we benchmark our method against state-of-the-art results for
two simple but highly inhomogeneous situations: finite-thickness jellium
slabs and the He atom. The results suggest that our scheme, while based on
approximations known to work well in uniform electron gases, also gives a good
treatment of strong inhomogeneity in one to three space dimensions. Details
follow.

The groundstate energy of an inhomogeneous many electron system with
external potential $v^{ext}(\vec{r})$ and groundstate electron density $n(%
\vec{r})\,$is given exactly by the constant-density adiabatic connection
formula of formal Kohn-Sham density functional theory: \cite
{AdiabatConnLangrethPerdew,AdiabaticConnGunnLundqvist} 
\begin{eqnarray}
E_{0} &=&T_{s}[n]+\int n(\vec{r})v^{ext}(\vec{r})d\vec{r}  \label{EGS} \\
&&+\frac{1}{2}\int \frac{e^{2}}{\left| \vec{r}-\vec{r}\,^{\prime }\right| }n(%
\vec{r})n(\vec{r}\,^{\prime })d\vec{r}d\vec{r}^{\prime }+E_{xc}[n]
\end{eqnarray}
\begin{equation}
E_{xc}[n]=\frac{1}{2}\int_{0}^{1}d\lambda \int \frac{e^{2}}{\left| \vec{r}-%
\vec{r}\,^{\prime }\right| }(n_{2\lambda }(\vec{r},\vec{r}^{\prime })-n(\vec{%
r})n(\vec{r}\,^{\prime }))d\vec{r}\,d\vec{r}\,^{\prime }.  \label{ExcACF}
\end{equation}
Here $T_{s}[n]=\hbar ^{2}(2m)^{-1}\sum_{k\,occ}\int \left| \nabla \phi _{k}(%
\vec{r})\right| ^{2}d\vec{r}$ $\,$is the KS kinetic energy, and $\{\phi
_{k}\}\,$are the occupied KS orbitals, eigenfunctions of the one-electron KS
potential $v_{KS}(\vec{r})$. $v_{KS}$ is defined to be such that {\em %
independent} electrons moving in $v_{KS}$ yield the true groundstate
density: 
\begin{equation}
v_{KS}\Longrightarrow \{\phi _{k}(\vec{r})\}:n(\vec{r})=\sum_{k\,occ}\left|
\phi _{k}(\vec{r})\right| ^{2}.  \label{DefVKS}
\end{equation}
 The $\lambda $ integration in (\ref{ExcACF}) accounts for the kinetic
part of the KS correlation energy. The groundstate pair distribution $%
n_{2\lambda }(\vec{r},r\,^{\prime })\,$is that of a ''$\lambda $-{\em %
system''}defined to have a reduced electron-electron interaction $\lambda
e^{2}/r_{12}$, and a modified external potential $v_{\lambda }^{ext}(\vec{r}%
)\,$chosen to maintain the true ($\lambda =1$) groundstate density at any $%
\lambda $: $n_{\lambda }(\vec{r})=n_{\lambda =1}(\vec{r})\equiv n(\vec{r})$.
Remarkably \cite{AdiabatConnLangrethPerdew}$,$\cite
{AdiabaticConnGunnLundqvist}, only the true external potential $%
v^{ext}\equiv v_{\lambda =1}^{ext}$ appears in (\ref{EGS}). Note that, from (%
\ref{DefVKS}), the KS potential of each $\lambda $-system is the same as
that of the true ($\lambda =1$) system, because the density is the same: 
\begin{equation}
v_{KS,\lambda }\equiv v_{KS}.  \label{VKSIndepLambda}
\end{equation}

The groundstate pair distribution $n_{2\lambda }$ in (\ref{ExcACF}) can be
related to the Kubo density-density response function $\chi _{\lambda }\,$of
the $\lambda $-system by the $T=0K$ fluctuation-dissipation theorem \cite
{LandauLifshitzStatPhys}, \cite{JDIntroTDDFTBneProc}, 
\begin{eqnarray}
&&n(\vec{r})n(\vec{r}\,^{\prime })(g_{\lambda }(\vec{r},\vec{r}\,^{\prime
})-1)\equiv n_{2\lambda }(\vec{r},\vec{r}\,^{\prime })-n(\vec{r})n(\vec{r}%
\,^{\prime })  \nonumber \\
&=&-\frac{\hbar }{\pi }\int_{0}^{\infty }\chi _{\lambda }(\vec{r},\vec{r}%
\,^{\prime },\omega =iu)du-n(\vec{r})\delta (\vec{r}-\vec{r}\,^{\prime }).
\label{Fluct-Diss}
\end{eqnarray}
Eq. (\ref{Fluct-Diss}) also introduces the equilibrium pair correlation
factor $g_{\lambda }(\vec{r},\vec{r}\,^{\prime })$.

The Kohn-Sham density-density response $\chi _{KS}$ \cite{GrossKohnPRL85} is
defined to be that of {\em independent} electrons moving in the KS\
potential $v_{KS}$. Note that, by (\ref{VKSIndepLambda}), 
\begin{equation}
\chi _{KS,\lambda }\equiv \chi _{KS}\equiv \chi _{\lambda =0}.
\label{ChiKSIndepLambda}
\end{equation}
$\chi _{KS}$ is exactly expressible \cite{GrossKohnPRL85} by perturbation
theory in terms of the KS orbitals $\{\phi _{k}\}.$ From this expression it
is readily shown that, when $\chi _{\lambda }$ is replaced by $\chi
_{\lambda =0}\equiv \chi _{KS}$, Eq. (\ref{ACFFDT}) gives the ``exact DFT
exchange'' energy, i.e. it gives the Hartree-Fock energy integral in which
the self-consistent Hartree-Fock orbitals are replaced by the KS orbitals $%
\phi _{k}$. Thus this formalism easily deals with exchange. Subtracting this
DFT exchange energy from (\ref{Fluct-Diss}) we obtain the exact DFT
correlation energy 
\begin{eqnarray}
E_{c} &=&\frac{-\hbar }{2\pi }\int d\vec{r}d\vec{r}\,^{\prime }\frac{e^{2}}{%
\left| \vec{r}-\vec{r}\,^{\prime }\right| }  \nonumber \\
\times \int_{0}^{1}d\lambda \int_{0}^{\infty }du[\chi _{\lambda }(\vec{r},%
\vec{r}\,^{\prime },\omega &=&iu)-\chi _{\lambda =0}(\vec{r},\vec{r}%
\,^{\prime },\omega =iu)].  \label{ACFFDT}
\end{eqnarray}
At each $\lambda \,$value the interacting and KS\ response are related
exactly by a Dyson-like screening integral equation\cite{GrossKohnPRL85} $%
\chi _{\lambda }=\chi _{\lambda =0}+\chi _{\lambda =0}*(\lambda
V_{coulomb}+f_{xc\lambda })*\chi _{\lambda },$where spatial convolution is
represented by a star. The xc kernel $f_{xc\lambda }\,$contains the
many-body xc effects and has tradionally been treated by a local density
approximation\cite{GrossKohnPRL85}\cite{VignaleKohnPRL96}\cite{EnOptFxc}.
Here, however, instead of using a local uniform-gas-based approximation we
effectively generate a nolocal $f_{xc}\,$self-consistently for the
particular inhomogeneous system. To do this we extend to nonuniform systems
the semiclassical approach of STLS. Thus we relate the independent-electron
and interacting responses by solving the time evolution equation (first
BBGKY hierarchy equation\cite{BoydSandersonPlasmaDynamics}) for the
one-electron distribution function $f(\vec{r},\vec{p},t)$ of the classical $%
\lambda $-system: 
\begin{eqnarray}
&&\left( \frac{\partial }{\partial t}+m^{-1}\vec{p}.\frac{\partial }{%
\partial \vec{r}}+\vec{F}_{\lambda }^{ext}(\vec{r},t).\frac{\partial }{%
\partial \vec{p}}\right) f(\vec{r},\vec{p},t)  \nonumber \\
&=&\int (\frac{\partial }{\partial \vec{r}}\frac{\lambda e^{2}}{\left| \vec{r%
}-\vec{r}\,^{\prime }\right| })\cdot \frac{\partial }{\partial \vec{p}}%
f_{\lambda }^{(2)}(\vec{r},\vec{p};\vec{r}\,^{\prime },\vec{p}\,^{\prime
},t)d\vec{r}\,^{\prime }d\vec{p}\,^{\prime }.  \label{Exact1BodyKinEq}
\end{eqnarray}
This equation is exact but requires the dynamic pair distribution $%
f_{\lambda }^{(2)}.$ The essential contribution of STLS\ was to use the
equilibrium pair-density factor $g_{\lambda }(\vec{r},\vec{r}\,^{\prime })\,$%
of Eq. (\ref{Fluct-Diss}) in a semi-classical truncation scheme 
\begin{equation}
f_{\lambda }^{(2)}(\vec{r},\vec{p};\vec{r}\,^{\prime },\vec{p}\,^{\prime
},t)\approx g_{\lambda }(\vec{r},\vec{r}\,^{\prime })f(\vec{r},\vec{p},t)f(%
\vec{r}\,^{\prime },\vec{p}\,^{\prime },t),  \label{STLSFactorization}
\end{equation}
where the true dynamic correlation factor $g_\lambda$ should depend on both the
momenta and the time, but this dependence is ignored and $g_\lambda$ is taken to
be the static, momentum-independent equilibrium density correlating factor from
Eq. (\ref{Fluct-Diss}). Using (\ref{STLSFactorization}) in (\ref
{Exact1BodyKinEq}) and linearizing about the equilibrium distribution, $f=$ $%
f_{0}(\vec{r},\vec{p})+\delta f(\vec{r},\vec{p},t)$, we obtain a closed
one-body kinetic equation 
\begin{eqnarray}
&&\left( \frac{\partial }{\partial t}+m^{-1}\vec{p}.\frac{\partial }{%
\partial \vec{r}}+\vec{F}^{(0)}(\vec{r}).\frac{\partial }{\partial \vec{p}}%
.\right) \delta f(\vec{r},\vec{p},t)  \nonumber \\
&=&-\delta \vec{F}^{eff}(\vec{r},t).\frac{\partial f_{0}(\vec{r},\vec{p})}{%
\partial \vec{p}}.  \label{LinSTLSKinEqu}
\end{eqnarray}
Here 
\[
\vec{F}^{(0)}(\vec{r})=\vec{F}_{0\lambda }^{ext}(\vec{r})-\int (\frac{%
\partial }{\partial \vec{r}}\frac{\lambda e^{2}}{\left| \vec{r}-\vec{r}%
\,^{\prime }\right| })g_{\lambda }(\vec{r},\vec{r}\,^{\prime })n_{0}(\vec{r}%
\,^{\prime })d\vec{r}\,^{\prime } 
\]
corresponds to the gradient of the KS potential in the quantal case, and is
independent of $\lambda \,$by choice of $F_{0\lambda }^{ext}(\vec{r})$.
Further, 
\begin{eqnarray}
\delta \vec{F}^{eff}(\vec{r},t) &=&\delta \vec{F}^{ext}(r,t)+\int \vec{W}%
_{\lambda }(\vec{r},\vec{r}\,^{\prime })\delta n(\vec{r}\,^{\prime },t)d\vec{%
r}\,^{\prime },  \label{DynScreenSTLS} \\
\;\;\vec{W}_{\lambda }(\vec{r},\vec{r}\,^{\prime }) &=&g_{\lambda }(\vec{r},%
\vec{r}\,^{\prime })\frac{-\partial }{\partial \vec{r}}\frac{\lambda e^{2}}{%
\left| \vec{r}-\vec{r}\,^{\prime }\right| }.  \label{EffForceW}
\end{eqnarray}
Because (\ref{LinSTLSKinEqu}) is linear and time invariant, its solution $%
\delta n(\vec{r},t)\equiv \int \delta f(\vec{r},\vec{p},t)d\vec{p}$ can be
expressed in the form 
\begin{equation}
\delta n(\vec{r},t)=\int \vec{\nu}_{0}(\vec{r},\vec{r}\,^{\prime
},t-t^{\prime }).\delta \vec{F}^{eff}(\vec{r}\,^{\prime },t^{\prime })d\vec{r%
}\,^{\prime }dt^{\prime }  \label{LinSolnNu0}
\end{equation}
where a $\vec{\nu}_{0}\,$is a $\lambda $-independent classical vector
response function giving the independent-electron density response to an
applied force, with $(\partial /\partial \vec{r}\,^{\prime }).\vec{\nu}%
_{0}=\chi _{\lambda =0}$.

In the case of{\em \ homogeneous} electron gases, $\vec{F}_{0}\,$is zero and 
$g_{\lambda }$ is a function only of the separation $R\equiv \left| \vec{r}-%
\vec{r}\,^{\prime }\right| $, and then $\vec{\nabla}\times \vec{W}=0$
\thinspace so that the effective pair force $\vec{W}_{\lambda }$ is
irrotational and can be expressed as a gradient of a scalar potential, $\vec{%
W}_{\lambda }(\vec{R})=-(\partial /\partial \vec{R})w_{\lambda }(R)$. Then,
assuming $\delta \vec{F}^{ext}$ comes from a potential $\delta V^{ext}$ we
can use integration by parts (Green's theorem) followed by space Fourier
transformation to write (\ref{LinSolnNu0}) in $q-$space as $\delta n=\chi
_{\lambda =0}\delta V^{eff}=\chi _{\lambda =0}(\delta V^{ext}
+w_{\lambda} \delta n)$.
This yields $\chi _{\lambda }(q,\omega )=\chi _{\lambda =0}(q,\omega
)(1-w_{\lambda }(q)\chi _{\lambda =0}(q,\omega ))^{-1}$. This equation
resembles a classical Random Phase Approximation (RPA), with $w_{\lambda
}(q) $ replacing the bare coulomb pair potential $4\pi \lambda e^{2}/q^{2}$.
 This response $\chi _{\lambda }$ depends on $g_{\lambda }$ via $w_{\lambda }$,
and $g_{\lambda }\,$is determined by $\chi _{\lambda }\,$via (\ref
{Fluct-Diss}), giving a closed selfconsistent scheme. $\,$STLS applied this
theory to the degenerate electron gas by replacing the classical
Boltzmann-equation density response $\chi _{\lambda =0}$ with the quantal
Lindhard response. Despite the crudeness of the factorization (\ref
{STLSFactorization}), the STLS\ formalism gives excellent correlation
energies for both 3D and 2D homogeneous electron gases, up to relatively
large values of the inter-electron spacing parameter $r_{s}$. For example,
in 3D $E_{c}^{STLS}\,$is within about 1\% of the 3D diffusion Monte Carlo
results\cite{CeperleyAlder} for $2<r_{s}\le 5$. The error is still under $%
4\% $ at $r_{s}=20\,$and under $7\%$ at $r_{s}=50$, a regime including gases
generally regarded as significantly correlated. These results are
significantly better than, e.g. recent GW-based many-body methods \cite
{vdWQuasi2DFromGWGodby}
 which give $20\%\,$%
error at $r_{s}=20$. The homogeneous STLS scheme has a number of
shortcomings including unphysical negative values of the on-top pair factor $%
g(\vec{r},\vec{r})$ and failure to satisfy the compressibility sum rule.
Further work addressing these difficulties\cite{ElCorrMetDensHayashiIV}, 
\cite{STLS4},\cite{ElCorrMetDensHasegawaII}, \cite{HolasRahmanQuantumSTLS87}%
, \cite{VashishtaSingwiElCorMetDens72} did not, however, significantly
improve the predicted uniform egas correlation energy. Therefore in the
present work we have concentrated on generalizing the original
semi-classical STLS\ scheme to inhomogeneous systems. This does not appear
to have been attempted previously:\ bilayered electron gases have certainly
been treated\cite{STLSCorrBilayerSzymSwNeilson94}, but these are isomorphic
to a two-species {\em homogeneous }2D electron gas. We will show that the
formalism is tractable for cases of genuine inhomogeneity.

In an {\em inhomogeneous }system we have$\vec{\nabla}\times \vec{W}_{\lambda
}=-\vec{\nabla}g_{\lambda }(\vec{r},\vec{r}\,^{\prime })\times \vec{\nabla}%
(\lambda e^{2}/\left| \vec{r}-\vec{r}^{\prime }\right| )\ne \vec{0}$ so that
there is no scalar potential corresponding to $\vec{W}_{\lambda }$, and the 
{\em vector} bare response $\vec{\nu}_{0}$ from Eq. (\ref{LinSolnNu0}) must
be used:\ the scalar version $\chi _{0}$ is not sufficient. This is an
essential difference between the inhomogeneous case and the homogeneous one.
As in the homogeneous case, we postulate that a degenerate Fermi system can
be treated via the above semi-classical analysis by using the quantal-Fermi
independent-electron response for $\vec{\nu}_{0}:\,$this also amounts to
using the quantal KS\ potential in place of its classical counterpart $\vec{F%
}_{0}$. By perturbation of the occupied independent-electron (Kohn-Sham, KS)
orbitals $\phi _{j}(\vec{r})\,$, we obtained for the inhomogeneous quantal
response at imaginary frequency $iu$%
\begin{eqnarray}
\vec{\nu}_{0}(\vec{r},\vec{r}\,^{\prime },\omega &=&iu)=\frac{1}{u}%
\mathop{\rm Re}%
[\frac{i\hbar }{m}\sum_{j}f_{j}\phi _{j}^{*}(\vec{r})  \nonumber \\
\times [G(\vec{r},\vec{r}\,^{\prime },E &=&\hbar \omega _{j}+i\hbar u)\vec{%
\nabla}\,^{\prime }\phi _{j}(\vec{r}\,^{\prime })  \nonumber \\
-\phi _{j}(\vec{r}\,^{\prime })\vec{\nabla}\,^{\prime }G(\vec{r},\vec{r}%
\,^{\prime },E &=&\hbar \omega _{j}+i\hbar u)]  \label{Nu0FromG}
\end{eqnarray}
where $f_{j}\,$is the Fermi occupation factor and $G$ is the Green function
for a single electron moving in the groundstate Kohn-Sham potential $%
v_{KS}(r)$. The Coulomb screening conditions (\ref{DynScreenSTLS}), (\ref
{EffForceW}), (\ref{LinSolnNu0}) for the inhomogeneous case can be written
as a Dyson-like ``screening'' integral equation for the interacting response 
$\chi_{\lambda}$:

\begin{eqnarray}
\chi_\lambda (\vec{r},\vec{r}^{\prime },\omega ) &=&\chi _{KS}(\vec{r},r^{\prime
},\omega )+
\int Q_\lambda (\vec{r},\vec{r}^{\prime \prime },\omega )
\chi_\lambda  (\vec{r}%
^{\prime \prime },\vec{r}^{\prime },\omega )dr^{\prime \prime } \nonumber ,\\
\label{STLSInhomScreening} \\
Q_\lambda (\vec{r},\vec{r}^{\prime \prime },\omega ) &=&\int 
\vec{\nu}_{0}(\vec{r},%
\vec{r}^{\prime \prime \prime }\omega ).\vec{W}_\lambda (\vec{r}^{\prime \prime
\prime },\vec{r}^{\prime \prime })d\vec{r}^{\prime \prime \prime }, \,\,
\chi_{KS} = \vec{\nabla}^{\prime}.\vec{\nu}_0 \nonumber \\  \label{DefQ}
\end{eqnarray}

We term this the ``inhomogeneous STLS" (ISTLS) scheme. To demonstrate its
feasibility and accuracy we have carried it out numerically for two highly
inhomogeneous but spatially symmetric cases, namely (i) charge-neutral
jellium slabs and (ii) a helium atom.

\begin{figure}[h]
\centerline{\psfig{file=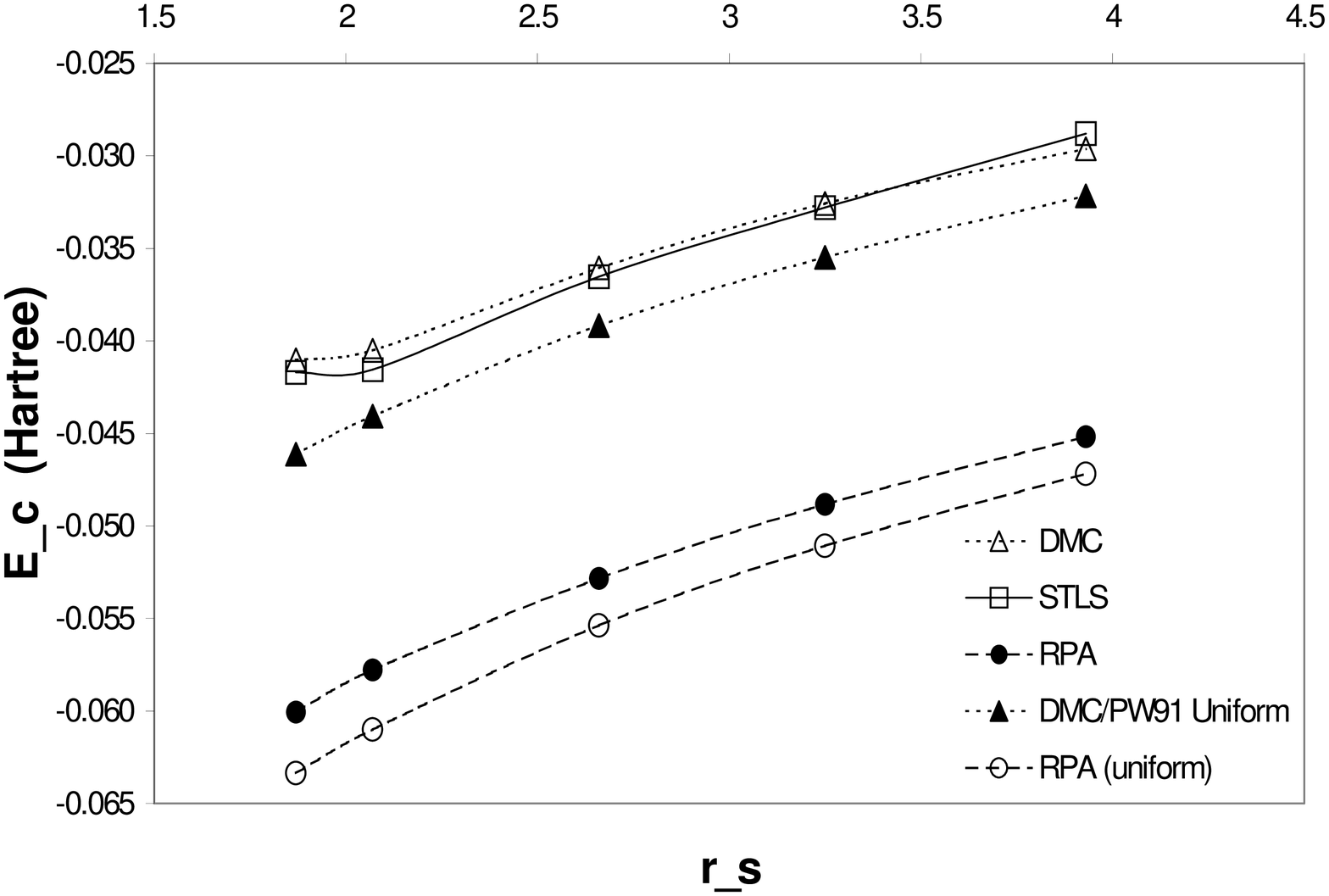,height=7cm,width=9cm}}
\caption{Correlation energy of neutral jellium slabs (Hartree/el)}
\end{figure}

The jellium slabs were first solved in the LDA (LDA-Perdew-Wang 1991\cite
{AccSimplEgasEcPerdewWangPRB92}) to give the selfconsistent groundstate
Kohn-Sham potential $v_{KS}^{LDA}(z)\,$and density $n(z)$, where $z$ $\,$is
the space coordinate in the thin dimension of the slab. 
Our ISTLS formalism was applied as a ``post-functional'' giving the
correlation energy starting from the fixed $v_{KS}^{LDA}(z)\,$, though of
course ideally one would choose a$\,v(z)\,$to minimize the total energy
including the STLS corrections. (This OPM method will also give an improved density $n({\vec r})$). 
Eqs. (\ref{Fluct-Diss}), (\ref{EffForceW}) and (\ref{Nu0FromG}) - (\ref{DefQ}%
) were then implemented. To perform the iterative refinement of the static
pair factor $g_{\lambda}(r,r^{\prime })\,$, we started with a $g$ of 
Hartree-Fock form
based on the occupied KS orbitals from the LDA groundstate. Finally (\ref
{ACFFDT}) was used.

Fig. 1 gives the slab correlation energy per electron (open symbols) for a
number of positive background densities $n_{0+}$, parametrized by the
dimensionless interelectron spacing $r_{s}=me^{2}\hbar ^{-2}(3n_{0+}/4\pi
)^{1/3}$. We show results from ISTLS (solid line), DMC \cite
{AcioliCeperleyJellSrfEnDMC} (dotted line) and RPA (dashed line) schemes,
the last obtained by setting the pair correlation factor $g$ to unity in
Eqs. (\ref{EffForceW}) and (\ref{Nu0FromG}) - (\ref{DefQ}). The thickness of
the positive background in each slab is $%
L=7.21r_{s}a_{B}$, to match the available DMC results. Results per electron
in the unbounded uniform gas are also shown (closed symbols). Agreement of $%
E_{c}^{ISTLS}$ with the slab DMC\ data is good, within 3\% .
This is comparable to the agreement of STLS with DMC for
the uniform 3D gas with $2\le r_{s}\le 20$ .

For small finite systems such as atoms one needs a self-interaction
correction (SIC) in the starting KS potential and density. Otherwise (as for
example when one uses the simple LDA) unrealistic response functions are
obtained because the asymptotic $-e^{2}/r$ potential is missing in $v^{KS}$.
We solved helium using the Krieger-Li-Iafrate exchange-only description\cite
{KLIPRA93} of the atomic groundstate. This has the advantage of a common
potential $v^{KLI}\equiv v^{KS}$ for all orbitals. 
 The explicit spherical form of Eqs. (\ref{Nu0FromG}) - (\ref{DefQ}) involves
spherical harmonics but is somewhat cumbersome because of the vector
character of $\vec{\nu}_{0}$. We obtained a total ISTLS He correlation
energy of $-40.0$ milliHartree, within $5\%$ of the ``exact''
nonrelativistic value\cite{CorrEnrgyAtomsIonsChakravorty93} of $-42.0\,mH$.
Our result  is of ''chemical accuracy''.
The KLI starting potential is adequate: we re-ran our method
starting from the numerically exact He KS\ potential\cite
{UmrigarGHeIsoXcPotls}, obtaining $<<$ $1\%\,$change in $E_{c}$.

In summary, we have derived an inhomogeneous generalization (``ISTLS'') of
the rather successful STLS uniform-gas correlation energy formalism: see
Eqs. (\ref{EffForceW}), (\ref{Nu0FromG}), (\ref{DefQ}), (\ref
{STLSInhomScreening}), (\ref{Fluct-Diss}), and (\ref{ACFFDT}). We have shown
that ISTLS gives good groundstate correlation energies in some highly
inhomogeneous electronic systems. The scheme can also encompass finite
temperatures, and plasmon calculations. An advantage of the ISTLS\ scheme is
that its pair correlation physics is ``self-tailored'' to the system at
hand, rather than ``stolen'' from some reference system as in some density
functional schemes, or guessed then optimized as in variational schemes. It
does this, moreover, without the need for solving explicit two-body dynamic
equations. We speculate that interesting density functionals for the
correlation energy might be derived by using the ISTLS scheme with
semi-local-density approximations for the bare response $\vec{\nu}_{0}$,
somewhat as in \cite{DobsonWangPRL99}. 

JFD acknowledges an Australian Research Council Large grant and the
hospitality of Prof. M. Combescot, Universit$\acute{e}$ de Paris VII. The
atomic application of ISTLS forms part of the PhD thesis of TG. We thank
Cyrus Umrigar for exact atomic $v_{KS}\,$data.

\vspace{-0.5cm}
\bibliographystyle{prsty}
\bibliography{1deg,2deg,3deg,dldf,gendft,graddft,hpt,hydro,jfdall,srfdynre,statsrf,teach,sic,data,vdw}

\end{document}